\documentclass[aps,prb,nofootinbib,twocolumn]{revtex4-2}
\usepackage{amssymb}
\usepackage{amsmath}
\usepackage{amstext}
\usepackage{amsfonts}
\usepackage[utf8]{inputenc}
\usepackage{graphicx}
\usepackage{slashed}
\usepackage{subfig}
\usepackage[colorlinks=true,allcolors=magenta]{hyperref}
\newcommand{\beq}{\begin{equation}}
\newcommand{\eeq}{\end{equation}}

\def\nn{\nonumber}

\def\be{\begin{eqnarray}}
	\def\ee{\end{eqnarray}}

\newcommand{\bea}{\begin{eqnarray}}
	\newcommand{\eea}{\end{eqnarray}}

\begin{document}

\title{  Dispersion Relations for Dislocation Modes and their Sensitivity to the  Lattice Structure
}

\author{Jin-Yun Lin}
\affiliation{Department of Physics, Carnegie Mellon University, Pittsburgh, PA 15213, USA}
\author{Shashin Pavaskar}
\affiliation{Department of Physics, Carnegie Mellon University, Pittsburgh, PA 15213, USA}
\affiliation{Department of Physics, University of Illinois at Urbana-Champaign, Urbana, IL 61801, USA}
\affiliation{IQUIST, University of Illinois at Urbana-Champaign, Urbana, IL 61801, USA}
\author{Ira Z. Rothstein}
\affiliation{Department of Physics, Carnegie Mellon University, Pittsburgh, PA 15213, USA}

\begin{abstract}
In this letter we show that the dispersion relation for the dynamical modes of dislocations (``dislons")
in solids is sensitive to the lattice symmetries. 
In particular, we show that in the IR, the dislon dispersion relation develops a logarithmic dependence on momenta for  approximately isotropic lattices whereas for non-isotropic lattices, the linear term dominates. 
The renormalization group flows for dislocation tension
are shown to be distinct for isotropic and anisotropic lattices.

\end{abstract} 

\maketitle

\section{Introduction}
	In this work, we explore the dynamics of dislocations in solids using an effective field theory approach. The physics of dislocations  is obviously quite mature  (see e.g.\cite{hirth_lothe}) but it  is perhaps fair to say that highly relevant pieces of the puzzle have  yet to be explored.	In particular, the dynamics of the excitations of the dislocation itself has only more recently begun to be investigated \cite{MLi,Li_new}\footnote{These works along with ours only consider straight dislocations which end on boundaries.}, though we believe that a first principles theoretical underpinning is still lacking, which is a hole we wish to fill herein. 
		
		The excitations of dislocations, which were termed ``dislons" in \cite{MLi,Li_new}, are  Goldstone bosons associated with the spontaneous breaking of space-time symmetries by the dislocation.
		When internal symmetries are broken, 
		the associated Goldstone bosons will have either a linear or quadratic
		dispersion relations, corresponding what what are called type I and type II respectively \footnote{More generally type I/II correspond to odd/even powers of the momentum. Furthermore, when rotational invariance is broken
			one can generate dispersion relations of the form $E= \sqrt{\sum_i c_i k_i^2}$.}.
		The number of type I and II Goldstones obeys $n_I+2 n_{II}= n_{BG}$ \cite{Nielsen:1975hm,Schafer,Watanabe:2012hr}, where $n_{BG}$ is the number of broken generators. 
		For the Goldstones arising from the breaking of  space-time symmetries \footnote{For a derivation of the inequality in the context of translational symmetry breaking, see \cite{Watanabe:2011dk}.} the situation becomes more
        complex. In  particular, Goldstones which propagate on sub-manifolds
        will in general have non-analytic dispersion relations.

		One would expect that the excitations of dislocations would correspond to canonical
		phonons which are type I.  However,  for embedded solids, the dispersion relation for these modes becomes non-analytic in the momenta. This non-analyticity results from integrating out gapless bulk phonon modes, analogous to what happens in the case of Kelvin waves in (super)-fluids \cite{alberto} or ripplons at super-fluid interfaces \cite{Watanabe:2014zza}.  In addition to this, the dispersion turns out to be sensitive to the boundary conditions of the solid. This is related to the fact for a one dimensional solid with open  ends, the tension will relax to zero, which leads to the vanishing of the
		quadratic spatial gradient term for the transverse modes  in the action. This well known fact, can be
		understood from the point of view of effective field theory as an example of
		a generic relaxation mechanism of a Wilson coefficients when the target space is bounded \cite{nicolis}. 
        What we will show here is that this same mechanism affects  the dislon dispersion relation for the transverse modes   such that it depends upon the nature of the symmetry of the underlying crystal lattice. This distinction will lead to phenomenological differences.

		To study the phenomenology of dislocations we need to understand
		the interaction with bulk phonons \footnote{It is also known that electron quasi-particle interactions with dislocations can play an important role in determining the superconducting temperature.
			This topic goes beyond the scope of this paper.
		}.
		However such interactions are non-local \footnote{In \cite{MLi,Li_new,zhang2016nonperturbative} this non-locality is manifest and is consequence of expanding around the stressed background of the dislocation. 
  However, expanding around a fixed background is limiting as discussed in the conclusions below.} since if we define
  the phonon,  $\pi^I(x)$, defined as the displacement of an atom away from 
		its equilibrium position, becomes multi-valued in the presence of a dislocation. This leads to an ambiguity as it is no longer clear  which atom ``belongs" to which position in the lattice.
		However, whichever choice is made to fix this ambiguity, the physics should remain
		unchanged, which is  nothing more than a gauge theory description of the physics  \cite{Kleinert:2008zzb}.
		For three dimensional solids, which we focus upon, dislocations are co-dimension two objects and, as such, couple topologically to anti-symmetric two form fields. Furthermore, 
		it is  known
		that derivatively coupled scalars, such as $\phi^I$,  have a dual description in terms of  an anti-symmetric two form gauge field \cite{townsend}, which couples locally to the dislocation.  This dual description of elasticity theory in three dimensions has been explored in \cite{zaanen},  which concentrated
		on the nature of the melting phase transition.  Here we will only be interested in the solid phase, though
		excitations of the dislocations modes we study here should be expected to play a roll in the transition as well.

		The goal of this paper is to develop a first principles theory for the dynamics of dislocations in solids. 
		In particular we will write down an effective field theory of the dislocation excitations which interact with the bulk phonons.
		A complete analysis of the non-linearities will be left for future work.

		\emph{Conventions} : We work with the mostly plus metric $(-,+,+,+)$, greek  (space-time) indices run $(0-3$) while Roman capital letters $I=1-3$
		correspond to the internal degrees of freedom (or, more precisely, the co-moving coordinate system). 
		Euclidean spatial indices are represented by small Roman letters.
		Note that after symmetry breaking, we can no longer
		distinguish between the capital and small Roman indices.
		We will also reserve the index $a$ to be in the directions transverse to the dislocation, i.e.
		$a=1,2$ while $i=1-3$.
		We work in units where $\hbar=c=1$.

\section{The String Action}

 To determine the action for the dislocation excitations we will build an effective
 field theory for an embedded solid string.
  Dislocations share commonalities  with   fundamental strings as well as with vortices in superfluids, with the important distinctions being the space-time and, in the case of vortices, internal symmetry breaking patterns. The effective field theory of vortices and their interactions in superfluids was worked out in the elegant paper \cite{alberto}, upon which we have leaned. There are some crucial distinctions however, as one might expect, between  vortices and dislocations, that distinguishes our analysis from \cite{alberto}.

We begin by considering the action of a physical/matter string in vacuum, and then will embed
the string into a solid. The  two scenarios will be distinguished by their space-time  symmetry breaking patterns.
A physical string is distinct from a fundamental string in that it lacks
reparameterization invariance (RPI) along the world sheet since we can now label
the matter elements along the string. Put another way, the matter string breaks  boost invariance along the directions of the string, and as such, the system includes an additional dynamical
degree of freedom corresponding to a longitudinal mode.
We choose to work in a covariant language, despite our eventual goal of calculating
in the non-relativistic limit, for ease of notation. 
Thus we will retain RPI and add an additional mode to the action which is given by
\beq
S= \int d^2 \sigma \sqrt{g} \: \mathcal{L}(K)
\eeq
where $g$ is the induced metric
\beq
g_{\alpha \beta}= \eta_{\mu \nu} \partial_\alpha X^\mu \partial_\beta X^\nu 
\eeq
and 
\beq
K= g^{\alpha \beta} \partial_\alpha \phi \partial_\beta \phi.
\eeq
This action contains three physical degrees of freedom, which by appropriate choice of
gauge corresponds to the transverse modes $X^a$ and $\phi$, and
is restricted to this form by the symmetries of the system, at leading order in the derivative expansion. For a quick review of the string action, one can refer to the appendix \ref{appendix:A}.

A dislocation is a string  embedded in a solid. It is a topological defect
which is characterized by a ``Burgers vector''  which is analogous to the vorticity
of a line defect in a superfluid and describes the holonomy of the embedding.
We will return to this point when we discuss the coupling of the dislocation to the phononic
bulk modes, to which we now turn.
\section{Effective field theory of solids}

A solid breaks space-time symmetries and simultaneously non-linearly realizes emergent
internal symmetries. The details of the symmetry breaking pattern are discussed in 
\cite{zoology,Nicolis:2013lma}. For our purposes it will suffice to recall
that all of the Ward identities can be saturated with just the three Goldstone bosons  $\pi^I$ associated with broken translations. 
The action for these Goldstones
can be fixed in a derivative expansion using a coset construction \cite{Pavaskar:2021pfo} or
via a Landau-Ginsburg construction \cite{Dubovsky_2006}, which we follow here.

  A solid is defined as a system for which the action is invariant under
  shifts of the co-moving coordinates, $ \phi^I \rightarrow \phi^I +a^I$. These shifts may be taken to be continuous
  in the long distance limit. Invariance under the internal $SO(3)$ rotations,
  will not however, be manifest for crystals, but only for ``Jelly like" objects. 
The action is then given by some arbitrary function of the quantity $B^{IJ}= \partial_\mu \phi^I \partial^\mu \phi^J.$ 
In the ground state we may align the co-moving coordinates with the
space-time coordinates, i.e. $\langle \phi^I \rangle =x^I$. Thus our power counting will be such that the first derivatives of the field $\phi^I$ can be of order one, and subsequent derivatives
will be suppressed. This vev breaks both the space-time and internal translations but preserves
the diagonal subgroup. For crystalline systems, there will be unbroken discrete rotations
which will play an important role below.

   With an eye on the fact that we will eventually be interested in taking the non-relativistic limit, we will follow \cite{Soper} and write the action in the following form
   \begin{equation}\label{scalarfieldaction}
   	S = -\int d^{D+1} x \hspace{0.1 cm} n( m+U(B^{IJ})),
   \end{equation}  
 where $n =\bar n \sqrt{\text{Det}(B^{IJ})}$ is the number density, $\bar n$ is the ground state number density, 
  $m$ is interpreted as the mass of the atoms sitting at the lattice sights and $U$ 
 is the internal energy functional.
 


 We may now consider excitations around the ground state via 
    \begin{equation}
    	\phi^I(\vec{x},t) = x^I + \pi^I(\vec{x},t)
    \end{equation}
    where $\pi^I(\vec{x},t)$ are the phonons. To obtain the action for the phonons, one can expand the function eq.(\ref{scalarfieldaction}) around the background $B^{IJ}=\delta^{IJ}$. At quadratic order this leaves


\begin{equation}\label{scalaraction1}
	\begin{split}
		S = \bar{n} \int d^4 x \hspace{0.1 cm}  &[\frac{1}{2}m \dot{\vec{\pi}}^2  -  \lambda_{IJ} (2 \partial^I \pi^J + \partial_{\mu}\pi^{I}\partial^{\mu} \pi^{J} ) - \\ & 2 \: \mathcal{C}^{IJKL} (\partial^I \pi^J  )(\partial^K \pi^L ) ]+....
	\end{split}
\end{equation}

$\bar n$ is the number density in the ground state and
we have Taylor expanded the functional ${U}(B^{IJ})= U(\delta_{IJ})+ \lambda_{IJ} (B^{IJ}- \langle B^{IJ} \rangle) +...$, 
where $\lambda_{IJ} \equiv \frac{\partial U}{\partial B_{IJ}}$
and  the elastic moduli tensor is given by  \beq C_{IJKL}\equiv \frac{\partial^2 U}{\partial B_{IJ} \partial B_{KL}}.\eeq 
More details on the non-relativistic limit of the action can be found in appendix \ref{appendix:B}.
The form of $\lambda_{IJ}$ and 	$C^{IJKL}$ depends upon the lattice symmetries. For a cubic lattice $\lambda_{IJ} \sim \delta_{IJ}$, but to achieve isotropy at rank four we need
icosahedral symmetry \cite{Kang:2018bqc}, in which case we would write
\begin{equation}
	\label{C}
	C^{IJKL} = c_2 \delta^{IJ}\delta^{KL} + c_3 (\delta^{IK}\delta^{JL} + \delta^{IL}\delta^{JK}).
\end{equation} 
For a general crystal one 
would need to decompose $C^{IJKL}$ into the proper invariant tensors for the crystal group of interest. Nonetheless one may have completely isotropic solids if they embody many crystalline domains or are glassy \cite{lubensky}, in which case dislocations may  no longer be  relevant. 


Notice that at this point the action depends upon both the symmetric as well as the anti-symmetric pieces of the phonon gradients $\partial_I \pi_J$ at quadratic order and above.
For a finite sample, under the assumption that any external applied stresses are isotropic, 
the anti-symmetric part can be set to zero at the cost of adding boundary terms \cite{nicolis}.

\section{Coupling of the Dislocation to bulk Phonons}


Up to this point, we have discussed the effective field theory for solids in terms of the co-moving coordinates $\phi^I(\vec{x},t)$ of the individual particles. However, as previously mentioned, the coupling of $\phi^I(\vec{x},t)$ to the dislocations is non-local in
this canonical description. Something similar happens in the case of superfluids, where it was shown  that the non-locality can be eliminated   by going to a dual (gauged) description \cite{townsend,alberto}.   In  3+1 solids, one may introduce anti-symmetric two-form fields $b_{\mu \nu}^I$ with $I=1,2,3$ to describe the low energy dynamics  \cite{zaanen} in such as way as to  localize
all couplings to the dislocation.

The idea \cite{townsend} is that when the theory is derivatively coupled we may change
variables from $\partial_\mu \phi^I$ to $F^I_\mu$ via a Legendre transformation.
Given a lagrangian  ${\cal L}(B^{IJ})$ we may define
\beq F_\mu^I=\frac{\delta   \mathcal{L}(B^{IJ} 
	)}{\delta  \partial^\mu \phi^I}, \eeq
The equations of motion for $\phi^I$ leads to the condition $\partial_\mu F^{\mu I}=0$,
which can be solved by defining
\beq
F_\mu^I= \epsilon_{\mu \nu \rho \sigma} \partial^\nu b^{\rho \sigma I}
\eeq
where $b^{\rho \sigma I}$ is the anti-symmetric vector valued two-form field. $F_\mu^I$ is
invariant under the gauge transformation
\beq
b_{\mu \nu }^I \rightarrow b_{\mu \nu} ^I+ \partial_\mu \xi_\nu^I-\partial_\nu \xi_\mu^I.
\eeq

Now using our action in \eqref{scalaraction1}, we find that at leading order, the dual fields are given by \footnote{Here we have utilized the  boundary terms to set the ``rotation" $\partial_i \pi_j \epsilon_{ijk}$ to zero. This also leads to a vanishing vev for the dual fields.}
\bea
\label{F1}
 F^{IJ} &=& -4\bar{n} \: \mathcal{C}^{IJKL} \partial^{K} \pi^{L} \nn \\
 F^{0I} &=& m\bar{n} \: \dot{\pi}^{I}    
\eea


We can decompose the two form  in terms of two vector fields $A_i^a$ and $B_i^a$ via  \cite{alberto}
\begin{equation}\label{perturb}
b_{jk}^{I}= \epsilon_{ijk}  B_{i}^I,  \qquad
b_{0k}^I=A_{k}^I.
\end{equation}
We then find 
\bea \label{dual2}
 F_A^I& =& \dot B_A^I -  (\vec \nabla \times \vec A)_A^I, \qquad F_0^I= \Vec{\nabla}\cdot\Vec{B}^I  \nn \\
\eea

Using \eqref{F1} and \eqref{dual2}, one obtains a local relation between the derivatives of the scalar and the dual fields $\vec A^I$ and $\vec B^I$, which can be used to derive the dual lagrangian \cite{zaanen}. We will gauge fix by choosing the Coulomb gauge, which  is not covariant, but since we will be interested in the non-relativistic case, this is of no consequence. To do this we include a term in the action 
\begin{equation}
S_{GF} = \frac{1}{2\zeta \bar n m} \int  d^4x (\partial^{i}b_{i\nu}^I)^2.
\end{equation}
In the limit $\zeta\rightarrow 0$  the fields $A$ and $B$ become transverse and longitudinal respectively, obeying  $\vec{\nabla}\cdot \vec{A}^I=0$ and $\vec \nabla\times \vec{B}^I=0$.  These correspond to nine conditions
that reduces the number of degrees of freedom (DOF) down to nine. The apparent
discrepancy with the required count of three DOF is clarified by examining the characteristic polynomial of the kernel, which contains three poles with quadratic dispersion relation. 

The leading order Lagrangian  in terms of these fields is then given by 
	\bea \label{bulkaction}
	m \bar n L^{LO}&=& \frac{1}{2\zeta}[(\partial_i A_{i}^I)^2+( \partial_i B_i^I)^2- ( \partial_i B_j^I)(\partial_i B_j^I)] \nn \\
	&+& \frac{1}{2 }(\partial_i B_{iI})^2 -\frac{m}{8 }C^{-1}_{KLIJ}( \dot B_{IJ}\dot B_{KL} \nn \\ 
	&+&  4(\vec \nabla \times \vec A)_{IJ}(\vec \nabla \times \vec A)_{KL} - 4 \dot B_{IJ} (\vec \nabla \times  \vec A)_{KL}) \nn \\
	\eea
Similarly one can derive the interactions in the bulk using the legendre transform, as shown in appendix \ref{appendix:D}.
As we have emphasized the central point of using the dual description is to construct a local effective action which captures the coupling of the phonons to the dislocation fields, which is encoded in the Kalb-Ramond action. The Kalb-Ramond action describes the leading coupling between the dislocation and the phonon gauge fields. 
The current associated with a $p$-dimensional defect is a $p+1$ form and hence a dislocation current in 3+1 dimensions is a two-form given by 
\begin{equation}\label{dislocation current}
	\begin{split}
		J^{\mu \nu}_I(x)  = \kappa^I  \int d\rho \: d\tau \hspace{0.1 cm}  \partial_{\tau} X^{[\mu} \partial_\rho X^{\nu ]} \hspace{0.1 cm} \delta^{(4)}(x-X(\tau,\rho)), 
	\end{split}
\end{equation}
where 
$\kappa^I$ is the Burgers vector.

 As is shown in appendix \ref{appendix:C}, the form of this current along with particle number conservation leads to the glide constraint, which states that dislocation motion in the direction orthogonal to the plane spanned by the Burgers vector and dislocation line is forbidden for an edge dislocation. The locality of this constraints implies that, for edge dislocation, there is only a single propagating dislon. 
 
 The Kalb-Ramond action takes the form
 \begin{equation}\label{KRcoupling}
S_{KR}   = \kappa^I \int d\tau \: d\rho \hspace{0.1 cm} \partial_\tau X^{\mu}  \partial_{\rho} X^{\nu} b^I_{\mu \nu} , 
\end{equation}
which is unique and gauge invariant upto a total derivative.
 One can notice that this term has no derivatives acting on the two-form  $b_{\mu \nu}^I$ which leads to a long range field. This should be contrasted with the scalar field theory for phonons where $\phi^I$  always comes with a derivative acting on it, 
 and the long range field can then only arise if the couplings are non-local as previously mentioned.

 
\section{The Dislocation Action}
	Let us  return to the action for a dislocation embedded in a solid. In this context, we now have at our disposal some bulk metric $G_{\mu \nu}$ in place of the Minkowski metric.	Thus we may define another world-sheet metric as
		\bea
		h_{\alpha \beta}= G_{\mu\nu} \partial_\alpha X^\mu  \partial_\beta X^\nu .
		\eea
		We now have many more invariants that we may write down.
		To be as general as possible we  may then consider actions of the
		form  \footnote{We have assumed that the volume element is controlled by the space-time metric. From the point of view of the EFT, we should also include a term with $\sqrt{-g}$ replaced by $\sqrt{-h}$. Doing so however, does not effect the results for the scaling derivatives.}
		\beq
		\label{S0}
		S= \int d^2\sigma \sqrt{-g} \: F(g^{\alpha \beta} \partial_\alpha \phi \partial_\beta \phi, h^{\alpha \beta} \partial_\alpha \phi \partial_\beta \phi) .
		\eeq
		In principle $F$ may depend upon contractions of $g$ and $h$ as well, as in the case of vortices \cite{alberto}, but these  terms will only contribute to non-linearities which are
		not of interest for this paper.
		
		To determine the dispersion relation for the transverse modes $X^a$, we choose the defect to lie on the z-axis
		and work in the gauge $X^0=\tau$, and $X^3=\rho=z$. The solid string coordinate
		will then be expanded around this configuration, that is $\phi= z$.
		Expanding out to quadratic order in the fluctuations  we have
      \bea
      \label{eq:act}
			S^{(0)} =  \int dt dz  \Bigg[-\frac{F}{2}\partial_t X^a \partial_t X^a +\frac{F-2F^\prime}{2}\partial_z X^a \partial_z X^a\nn\\
   - \dot F  \frac{ G_{ab}}{ G_{zz}^2}(\partial_z X^a)(\partial_z X^b) +2 \ddot F\Big( \frac{G_{az}}{G_{zz}}\partial_z X^a\Big)^2- \dot F(\frac{G_{az}\partial_t X^a}{G_{zz}})^2 \Bigg]\nn\\
		\eea
where dots/primes represent derivatives with respect to the second/first argument. The stress along the dislocation axis is given by
\beq
\label{eq:tens}
T_{zz}= 2F^\prime-F+2 \frac{\dot F}{G_{zz}}.
\eeq

We must impose some set of boundary conditions for the solid. It is most
natural/typical to choose the ends to be free, in which case the
tension, of the solid, as well as the dislocation relaxes to zero \cite{nicolis}.
Indeed we can see that for a solid with cubic symmetry, where the 
components of $G$ are diagonal, the vanishing of the tension also implies
that that terms quadratic in momenta vanish which corresponds to the
well known fact that a bar with open ends has a quadratic dispersion
relation $\omega \sim k^2$ for its sound modes as opposed to the canonical
linear dispersion. This relaxation is also responsible for the independence of the bulk action from the anti-symmetric part of the gradient $\partial_I \pi_J$, which is sometimes called the ``angulon" (see \cite{nicolis} for a discussion in the context of field theory).
		
		A particularly interesting aspect of dislons is that they  mix with the bulk modes, as can be seen from re-writing the Kalb-Ramond action in terms of the vector components of the two-form fields in our gauge $X^0=\tau = t$ and $X^3=\rho=z$, 
\begin{equation}\label{KR}
\begin{split}
S_{KR} &=  \kappa^I \int dt \: dz \hspace{0.1 cm}  \partial_z X^{k} A_{k}^I  +  \partial_t X^{k} \partial_z X^i  \epsilon_{kij}B^I_j
\end{split}
\end{equation}
 
This mixing implies that the dispersion relation will not be polynomial in
derivatives anymore.
Since we are interested in the behavior of transverse dislon modes, let us compute these corrections using the above action. 
\bea
\label{eq:correctionPreInt}
S^{(1)} = \frac{\kappa^I \kappa^J}{2}  \int \frac{d\omega}{2\pi} \frac{d^3 k}{(2\pi)^3}  X^a_{-\omega,-k_z} \:  [\epsilon_{ac} \epsilon_{bd} \: (G_{B}(k))_{cd}^{IJ}  \: \omega^2 \nn \\ + (G_{A}(k))_{ab}^{IJ} \: k_z^2 + 2\omega k_z\epsilon_{bc}(G_{AB}(k))_{ac}^{IJ}  ] X^b_{\omega,k_z} \nn\\
\eea
where $G_A(k)$, $G_B(k)$ and $G_{AB}(k)$ are the correlators for the dual fields. We refer to appendix \ref{appendix:D} for the explicit form of the dual propagators for an isotropic solid. Integrating over the transverse momenta will induce a logarithmic running of the coefficients in eq. \eqref{eq:act}.
In the case of an isotropic solid, the phonon corrections contribute only to the diagonal terms in the dislon action. 
In terms of the bare parameters which we now label with subscript $B$, the effective action in momentum space for the screw ($\kappa_i=\delta_{i3}\kappa$) and edge ($\kappa^i=\delta^{i1}\kappa$) dislocations are given by 
 \begin{widetext}
 \bea\label{eq:screw}
    S_{screw} &=& m \bar n\int \frac{d \omega}{2 \pi} \int \frac{d k}{2 \pi} \sum_{a=1,2}X_{-\omega,-k}^a\Big(-\frac{\omega^2}{2}\Big(\frac{1}{m \bar n}F_B +\frac{\kappa^2}{4\pi\epsilon}-\frac{\kappa^2}{4\pi}\ln\frac{k^2}{\mu^2}\Big)\nn\\&+&\frac{1}{2}k^2\Big(\frac{1}{m \bar n}(F_B-2 F_B^\prime-2\frac{\dot{F}_B}{G_{zz}})+\frac{1}{\pi}\frac{\kappa^2(c_L^2-c_T^2)c_T^2}{c_L^2\epsilon}-\frac{\kappa^2 c_T^2}{ \pi}\frac{(c_L^2-c_T^2)}{c_L^2}\ln\frac{k^2}{\mu^2}\Big)+\frac{\kappa^2c_T^4k^4}{2\pi\omega^2}\bigg(\frac{1}{4}\nn\\
    &-&(1-\frac{\omega^2}{c_L^2k^2})\ln(1-\frac{\omega^2}{c_L^2k^2})+(1-\frac{\omega^2}{c_T^2k^2})\ln(1-\frac{\omega^2}{c_T^2k^2})\bigg)
    +\frac{2\kappa^2c_T^6}{8\pi}\frac{k^6}{\omega^4}(1+\frac{\omega^6}{c_T^6k^6})\ln(1-\frac{\omega^2}{c_T^2k^2})\Big)X_{\omega, k}^a\nn\\
\eea
 \bea\label{eq:edge}
    S_{edge} &=& m \bar n\int \frac{d \omega}{2 \pi} \int \frac{d k}{2 \pi}  X_{-\omega,-k}^1\Big(-\frac{\omega^2}{2}\Big(\frac{1}{m\bar n}F_B+\frac{\kappa^2}{4\pi\epsilon}-\frac{\kappa^2}{4\pi}\ln\frac{k^2}{\mu^2}\Big)\nn\\&+&\frac{1}{2}k^2\Big(\frac{1}{m\bar n}\big(F_B-2F_B^\prime-2\frac{\dot{F}_B}{G_{zz}}\big)+\frac{3\kappa^2c_T^2}{16\pi\epsilon}-\frac{3\kappa^2c_T^2}{16\pi}\ln\frac{k^2}{\mu^2}\Big)
    +\frac{\kappa^2c_L^4k^4}{32\pi\omega^2}(1-\frac{\omega^2}{c_L^2k^2})^2\ln(1-\frac{\omega^2}{c_L^2k^2})\nn\\&-&\frac{3\kappa^2c_T^4k^4}{32\pi\omega^2}-\frac{\kappa^2c_T^2k^2}{32\pi\omega^4}(3c_T^4k^4+2c_T^2k^2\omega^2+3\omega^4)(1-\frac{\omega^2}{c_T^2k^2}) \ln(1-\frac{\omega^2}{c_T^2k^2})\Big)X_{\omega, k}^1\nn\\
\eea
\end{widetext}
We have also traded the Lam\'e parameters in the elastic moduli tensor for $c_L$ and $c_T$, which are the velocities for the longitudinal and transverse phonons respectively that obey the bounds $c_L^2>\frac{4}{3} c_T^2$ \cite{Landau:1986aog}.
There is only one transverse mode of the edge dislocation once the glide constraint is imposed (see appendix $C$).
Note the existence of a branch cut
corresponds
to the production of a physical phonon in the intermediate state. Above threshold $\omega>c_{T/L}k_z$,
the dislon is unstable to decay to phonons. It is interesting to note that at the branch point the log terms vanish for the edge but not the screw dislocation.
We currently do not have a concrete understanding of this.

 We define the renormalized parameters via the relations
\bea F_B&=&\mu^{2 \epsilon}(F_R(\mu)-\kappa^2\frac{m \bar n}{4\pi}\frac{1}{\epsilon})\nn \\ 
 \bar F_B&\equiv&(F_B-2F_B^\prime-2\frac{\dot{F}_B}{G_{zz}})\nn\\&=&\begin{cases}
     \mu^{2 \epsilon}(\bar F_R(\mu)- \kappa^2\frac{ m \bar n (c_L^2-c_T^2)c_T^2}{\pi c_L^2}\frac{1}{\epsilon})\hspace{24pt}\text{screw}\\
     \mu^{2 \epsilon}(\bar F_R(\mu)- \kappa^2 \frac{3 m\bar n c_T^2}{16 \pi }\frac{1}{\epsilon})\hspace{50pt}\text{edge}
 \end{cases}
\eea
Choosing the sample size $L$ as the relevant IR scale, the running couplings may be written as
\bea
F_R(\mu)&=&F_R(1/L)- \kappa^2\frac{ m \bar n}{4\pi} \ln(\mu^2 L^2) \nn \\
\bar F_R(\mu)&=& \begin{cases}
    \bar F_R(1/L)- \kappa^2 \frac{m\bar n(c_L^2-c_T^2)c_T^2}{\pi c_L^2}\ln(\mu^2 L^2)\hspace{5pt}\text{screw} \\
    \bar F_R(1/L)- \kappa^2 \frac{3 m \bar n c_T^2}{16\pi }\ln(\mu^2 L^2) \hspace{32pt}\text{edge}
\end{cases}
\eea
From this result we can see that the tension eq.  (\ref{eq:tens}) is renormalized just as in the case of the superfluid vortex \cite{alberto}. A measurement of the tension must be made in order to fix it at some scale.
Given our assumption of open boundary conditions, it is only sensible
to impose that $T_{zz}(\mu=1/L)=0$, where $L$ is the sample dimensions.
Thus given, our open boundary conditions implies that $T_{zz}(1/L)=0$ we can predict the tension at any given scale,
\beq
T_{zz}(\mu)=\begin{cases}
 \kappa^2\frac{m \bar n (c_L^2-c_T^2)c_T^2}{\pi c_L^2}\ln(\mu^2 L^2)\hspace{45pt}\text{screw} \\
\kappa^2\frac{3 m \bar n c_T^2}{16 \pi}\ln(\mu^2 L^2)\hspace{72pt}\text{edge}
\end{cases}
\eeq

\section{Low Frequency Approximation}
The effective action computed in \eqref{eq:screw} and \eqref{eq:edge} is well-defined in the low frequency limit, $\omega \rightarrow 0$. 
Assuming that $\omega$ is approaching zero faster than $k$, which
we will check aposteriori, 
and expanding the action in $\omega$, one obtains  
\begin{widetext}
    \bea \label{eq:screwaction}
    S_{screw} = m \bar n \int \frac{d \omega}{2 \pi} \int \frac{d k}{2 \pi} \sum_{a=1,2} X_{-\omega,-k}^a\Big(-\frac{\omega^2}{2}\Big(\frac{1}{m\bar n}F_R(1/L)-\frac{\kappa^2}{4\pi}\log(k^2 L^2)\Big) -\frac{\kappa^2 c_T^2}{2\pi}(1-\frac{c_T^2}{c_L^2})k^2\ln(k^2 L^2)\Big)X_{\omega, k}^a\nn\\
\eea
\bea \label{eq:edgeaction}
    S_{edge} = m \bar n \int \frac{d \omega}{2 \pi} \int \frac{d k}{2 \pi} X_{-\omega,-k}^1\Big(-\frac{\omega^2}{2}\Big(\frac{1}{m\bar n }F_R(1/L)-\frac{\kappa^2}{4\pi}\ln(k^2 L^2)\Big) -\frac{3\kappa^2c_T^2}{32\pi}k^2\ln(k^2 L^2)\Big)X_{\omega, k}^1\nn\\
\eea
\end{widetext}
We have redefined the renormalized parameters $F_R \rightarrow m \bar n \: F_R$ for simplicity. The dispersion for the modes can now be computed analytically and is given by
 \beq \label{dislondispersion}
 \omega^2=
 \frac{\gamma^2 \kappa^2 \: m \bar n }{\pi(-F_R(1/L)+\kappa^2\frac{m \bar n}{2\pi} \ln(kL))}c_T^2k^2 \ln(k L)
 \eeq
with 
\bea
\gamma = \begin{cases}
    \sqrt{2(1-\frac
{c_T^2}{c_L^2})} \hspace{0.8cm} \text{screw} \\ \sqrt{\frac{3}{8}}. \hspace{2cm} \text{edge}
\end{cases}
\eea
We see that the low frequency approximation is self-consistent since
$\omega$ is approaching zero faster than $k$, as $k$ approach its IR limit of $1/L.$

  One can see that the frequency gets a logarithmic dependence on the momenta. 
  The group velocity for the dislon modes is given by

  \bea
  \label{eq:groupV}
     v(k) = \frac{d\omega}{d k} = \kappa\frac{\gamma c_T} {\sqrt{\pi}}\sqrt{\frac{m \bar n \ln(kL)}{-F_R(1/L)+\kappa^2\frac{m\bar n}{4\pi} \ln(kL)}}\nn\\
       \Bigg[1- \frac{n^2}{8\pi}\frac{m\bar n }{-F_R(1/L)+\kappa^2\frac{m\bar n}{4\pi} \ln(kL)}+ \frac{1}{2\ln(kL)}\Bigg]\nn\\
  \eea


In this approximation we can also discuss what happens for an non-isotropic lattice. As in the isotropic case, the exchange of bulk modes renormalizes the couplings in \eqref{eq:act} and thus leads to a  running of the Wilson coefficients.
However, in the non-isotropic case, the relaxation conditions that sets $T_{zz}$ to zero (see eq. \ref{eq:tens}) no longer leads to a vanishing coefficient of the
$k^2$ term (i.e. with no log) in the action. Thus while at short distance $kL\gg 1$
the isotropic and an-isotropic cases will have the same dispersion relation, at
long distances they will differ.

\section{Conclusion}
In this paper, we have presented a first principles geometric construction of the effective field theory of dislocations in solids which is valid at distance
scales large compared to the dislocation width. While we have only studied the
quadratic part of the action in this paper, there is no obstruction to including
non-linearities at arbitrary orders in derivatives. 
We have emphasized the interesting fact that the dispersion relation for the excitations of dislocations called ``dislons" depends on the lattice structure. While the dispersions in both isotropic and anisotropic solids get logarithmic dependence on the momenta, the absence of a linear term in the isotropic case makes their dispersions differ at long-wavelengths.
As a result, the dislon velocities are also sensitive to the lattice structure.    
We have shown that $\lim_{k \rightarrow 1/L} v_{iso}/v_{aniso}\sim 1/\sqrt{\log(kL)}$ where $v_{iso}/v_{aniso}$ are the dislocation velocities for isotropic/anisotropic crystals with $L$ being the sample size. 

We also point out how our construction is distinct from the pioneering work on the field theory of dislons \cite{MLi,Li_new}. As a consequence of our geometric construction our action will automatically realize all of the non-linearly realized broken space-time symmetries, both on the dislocation as well as in the bulk. Thus we have the ability to calculate non-linear corrections at arbitrary order in the derivative expansion. It has not clear how the works \cite{MLi,Li_new} would incorporate such corrections. These papers correctly reproduce the log dependence in the dispersion relation, by  expanding about the expectation value of the static phonon field for a straight string. However, this method is limited in the sense that is most easily seen, by considering the force between two dislocations since the background will now change in a non-perturbative way. In our construction, since we work in the dual theory where the phonon coupling to the dislocation can be described in a local way, we can can calculate the force using the same action in a straight forward fashion.  Previous works on this subject did not utilize the boundary conditions which set the long distance tension to zero, and therefore did not address the difference in RG flow as one looses isotropy.

  Lastly, there has been considerable work on understanding phonon  \cite{zhang2016nonperturbative,lund2019scattering,lund2020scattering}  and electron transport \cite{MLi,Li_new,li2017tailoring,fu2017oscillatory} in the presence of dislocations. It would be interesting to carry out these computations in our formalism and we leave them for future work.
\section{Acknowledgements}
The authors would like to thank Riccardo Penco for many discussions on \cite{alberto}. We also benefited from discussions with Amit Acharya and Alberto Nicolis.
 IZR is supported by DOE grants DE- FG02-04ER41338 and FG02- 06ER41449.
\appendix
\section{The String Action}
\label{appendix:A}
Consider a fundamental string embedded in a $D$-dimensional spacetime. The string traces out a two-dimensional surface (worldsheet) as it evolves in time which is parameterized by $(\tau,\sigma)$. The action for a fundamental string is proportional to the area of the string worldsheet and is given by  
\begin{equation}\label{action}
\begin{split}
		S &= -T\int dA \\
		  &= -T \int d\tau \: d\sigma \: \sqrt{- \text{det} \: G_{\alpha \beta}} 
\end{split}	
\end{equation} 
where $G_{\alpha \beta}=g_{\mu \nu}\partial_{\alpha}X^{\mu} \partial_{\beta}X^{\nu}$ is the induced worldsheet metric, $g_{\mu \nu}$ is the space-time metric and $T$ is the tension in the string. $X^{\mu}(\tau,\sigma)$ describe the $D$-dimensional embedding coordinates of the string.
Here $(\mu,\nu)=0,1,...,D-1$ and $(\alpha,\beta)$ run over $(\tau,\sigma)$. This reduces to the standard Nambu-Goto action for relativistic strings in flat space-time.  

This action is invariant under re-parametrization of the time-like coordinate $\tau \rightarrow \tau'(\tau,\sigma)$, which is the consequence of introducing a redundant degree of freedom $X^0(\tau,\sigma)$. This results in a manifestly Lorentz invariant action \eqref{action}. In addition to this, the action is also invariant under the re-parametrization $\sigma \rightarrow \sigma'(\tau,\sigma)$. This can be understood from the fact that the boosts are unbroken along the string and hence the excitations of the string along itself $X^3(\tau,\sigma)$ are unphysical and one should always be able to gauge it away. One can choose the static gauge where $X^0=\tau$ and $X^3=\sigma$ such that the action is now only a function of the transverse coordinates $X^a(\tau,\sigma)$, where $a=(1,2)$. 

In the case of a solid string, boosts are now broken along the string and hence the action is no longer invariant under reparametrizations of $\sigma$. To restore such an invariance, one needs to introduce an additional degree of freedom $\phi(\tau,\sigma)$ which lives on the string worldsheet. This describes the co-moving coordinates of the elements of the solid string. 
Hence the action for solid string should now be invariant under additional constant shifts $\phi(\tau,\sigma) \rightarrow \phi(\tau,\sigma) + c $. Thus, at leading order in the derivatives the action for a solid string is an arbitrary function of $B = G^{\alpha \beta} \partial_{\alpha} \phi \partial_{\beta} \phi$. 
\begin{equation}
	S= \int d\tau \: d\sigma \: \sqrt{-\text{det} \: G_{\alpha \beta}} \: F(B)
\end{equation}
Once again choosing the static gauge, the only relevant degrees of freedom correspond to the transverse excitations $X^a(\tau,\sigma)$ as well as $\phi(\tau,\sigma)$. 

\section{Action for a Non-relativistic solid}
\label{appendix:B}
In Sec. III, we derived the quadratic action for phonons in a non-relativistic solid. The starting point of our action was
  \begin{equation}\label{scalarfieldaction}
   	S = -\int d^{D+1} x \hspace{0.1 cm} n( m+U(B^{IJ})),
   \end{equation}  
   The number density $n$ can be defined  via the current $J^{\mu}$ for the comoving coordinates
    \beq \label{scalarcurrent}
J^\mu = \frac{\bar{n}}{6}\epsilon^{\mu \nu \rho \sigma} \epsilon_{IJK} \partial_\nu \phi^I \partial_\rho\phi^J \partial_\sigma \phi^k
\eeq
where $\bar n$ is the ground state number density in the rest frame.
The current obeys
\beq
d * J=0,
\eeq
i.e. it is algebraically conserved (off-shell) and expresses the matter conservation.
We can write an invariant expression for the number density which reduces to det($B^{IJ}$) in the local rest frame
 \beq n=\sqrt{-J^{\mu}J_{\mu}}. \eeq 

 The corresponding velocity field is given by
\beq
\label{v}
u^\mu= J^\mu/\sqrt{-J^2}.
\eeq
such that $u^2=-1$ and in the local rest frame $u^0=1$.
We can thus rewrite the action in \eqref{scalarfieldaction} as 
\beq
S = \int d^4 x \hspace{0.1 cm}  -J^0 \sqrt{1-\vec{u}^2} \: ( m + U(B^{IJ}) ) 
\eeq

In the NR limit, one can expand the above action in $\vec{u}$ and take the $c\rightarrow \infty$ to obtain \beq
S = J^0 \int d^4 x \: \frac{m\vec{u}^2}{2}  - U(B^{IJ}) 
\eeq
In the above action we have dropped the contributions from the term $J^0m $ since it does not contribute to the equations of motion, which can be seen from the fact that the current $J^{\mu}$ can be written as a total derivative $J^{\mu}=\partial_{\nu}(J^{\nu}x^{\mu})$. Expanding the fields $\phi^I$ about its vev, one can obtain quadratic lagrangian for phonons
\begin{equation}\label{scalaraction}
	\begin{split}
		S = \bar{n} \int d^4 x \hspace{0.1 cm}  &[\frac{1}{2}m \dot{\vec{\pi}}^2  -  \lambda_{IJ} (2 \partial^I \pi^J + \partial_{\mu}\pi^{I}\partial^{\mu} \pi^{J} ) - \\ & 2 \: \mathcal{C}^{IJKL} (\partial^I \pi^J  )(\partial^K \pi^L ) ]+....
	\end{split}
\end{equation}
$C^{IJKL}$ is referred to as the elastic moduli tensor. The number of independent components in the elastic moduli tensor for a crystal depends on its crystal group. In the case of  isotropic solids, there are two independent components, and the tensor is the linear combination of the symmetric-traceless ($S_{IJKL}$) and trace ($T_{IJKL}$) projectors. In three spatial dimensions, the tensor is given
\bea 
C_{IJKL} &=& 3 \tilde\kappa \: T_{IJKL}+2\mu \:S_{IJKL}
\eea
with
\bea
S_{IJKL}  &=&\frac{1}{2}\left(\delta_{IK} \delta_{JL}+\delta_{IL} \delta_{JK}\right)-\frac{1}{3} \delta_{IJ} \delta_{KL} \nn\\
T_{IJKL}  &=&\frac{1}{3} \delta_{IJ} \delta_{KL}
\eea
where $\tilde \kappa$ and $\mu$ are the bulk and shear modulus respectively.
\section{Glide Constraint}
\label{appendix:C}
In a crystal, dislocation motion can lead to plastic deformation. Dislocations can slip under relatively low stress in the plane spanned by the dislocation line and its Burgers vector. On the other hand, the glide constraint states that motion perpendicular to the slip plane, called a climb motion, is prohibited. Given the definition of particle number current in the previous section, one can show that in linear elasticity, the glide constraint is a consequence of matter conservation. Writing out the matter conservation equation gives
\bea
 \label{B}
\partial_\mu J^\mu  = \frac{\bar{n}}{6}(\epsilon^{\mu\nu\rho\sigma}\partial_\mu\partial_\nu \phi^I)\partial_\rho\phi^J \partial_\sigma \phi^K\epsilon_{IJK}\eea
which vanishes unless there is a dislocation
which allows a violation of the integrability condition. Expanding $\phi^I$ about $x^I$, we obtain to next to lowest order the expression of the glide constraint
\bea\partial_\mu J^\mu  = \frac{\bar{n}}{6}\epsilon_{IJK}\tilde{J}^{I}_{JK}=0\eea
where
$\tilde{J}^{I}_{\mu\nu} = \epsilon^{\mu\nu\rho\sigma}\partial_\rho\partial_\sigma \phi^I$ is the dislocation current described in terms of the canonical phonon (see chapter two of \cite{Kleinert:2008zzb}). 
This current $\tilde{J}^{I}_{\mu\nu}$ is equivalent to the one given in  eq(\ref{dislocation current}),
which then implies that for the edge dislocation there is only
one tranverse dislon mode. 
In \cite{Cvetkovic_2006} it was argument that this constraint should not hold point by point, i.e. one needs to integrate it over the length of the string. This would imply that only the zero mode in that direction vanishes and that therefore that in principle there should be 
two dislon polarizations. This is the so-called
``leaky glide constraint". It is not clear to the authors if this implies the existence of a small gap or not. 

\section{Duality}
\label{appendix:D}
In this section, we provide details on the duality transformation in solids. 
Given an action for a derivatively coupled scalar ${\cal L}(K^{IJ})$ we may write its legendre transform as 
\beq\label{dualaction}
{\cal G}(Y^{IJ})= \partial^\mu \phi^I \frac{\delta   \mathcal{L}(K^{IJ} 
	)}{\delta  \partial^\mu \phi^J}- \mathcal{L}(K^{IJ} ),
\eeq
where $Y^{IJ}=F_\mu^I F_{\mu J}$ and \beq F_\mu^I=\frac{\delta   \mathcal{L}(K^{IJ} 
	)}{\delta  \partial^\mu \phi^J}, \eeq
	and the inverse is given by
\beq
\label{legend}
\mathcal{L}(K^{IJ} )= \frac{\delta {\mathcal G}(Y^{IJ})}{\delta F^{\mu I}} F^{\mu I }-\mathcal{G}(Y^{IJ})
\eeq
 \beq{} \partial_\mu \phi^I=\frac{\delta \mathcal{G}(Y^{IJ} 
	)}{\delta F^{\mu I}}. \eeq
For the Legendre transform, and its inverse, to be well defined $\mathcal{L}$ must be convex.
The equations of motion for $\phi^I$ leads to the condition $\partial_\mu F^{\mu I}=0$,
which can be solved by defining
\beq
F_\mu^I= \epsilon_{\mu \nu \rho \sigma} \partial^\nu b^{\rho \sigma I}
\eeq
and $b^{\rho \sigma I}$ is the anti-symmetric vector valued two form field.
As discussed in the main text, the presence of open boundary conditions implies that stress in the ground state should vanish. This is equivalent to setting $\lambda^{IJ}=0$ in \eqref{scalaraction}. Now using our action in \eqref{scalaraction}, we find that at linear order
\bea
\label{F}
 F^{IJ} &=& -4\bar{n} \mathcal{C}^{IJKL} \partial^{K} \pi^{L} \nn \\
 F^{0I} &=& \bar{n} m \dot{\pi}^{I}    
\eea

Thus to leading order we have
\bea\label{duality}
\partial^{K} \pi^{L}&=& - \frac{1}{4\bar{n}}\mathcal{C}^{(-1)KLIJ}  F^{IJ} \nn \\
\dot{\pi}^I &=& \frac{1}{\bar{n}m } F^{0I}.
\eea

Using the above relations, we may now write \eqref{dualaction} at leading order as
\beq
\mathcal{G^{LO}}= \frac{1}{\bar{n}m } ( F^{0I})^2- \frac{1}{4 \bar{n}}  C^{(-1)IJKL} F^{IJ}  F^{KL}- \mathcal{L}^{LO}(  F^{\mu I}).
\eeq
where 
\bea
\mathcal{L}^{LO}(  F_\mu^I)&= &\frac{1}{2\bar{n} m} ( F_0^I)^2 -\frac{1}{8\bar{n}}\mathcal{C}^{(-1)IJKL}  F^{IJ}  F^{KL} \nn \\
\eea
which gives the leading order dual action
\beq
\mathcal{G^{LO}}= \frac{1}{2\bar{n}m } ( F^{0I})^2- \frac{1}{8 \bar{n}}  C^{(-1)IJKL} F^{IJ}  F^{KL}.
\eeq


Since the action is written in terms of gauge invariant variables  it is invariant under the transformations $b_{\mu \nu}^I \rightarrow b_{\mu \nu}^I  +\partial_{\mu}\lambda_{\nu}^I - \partial_{\nu}\lambda_{\mu}^I$.   The gauge transformation parameter itself has a  further redundancy $\lambda_{\mu}\rightarrow \lambda_{\mu}+\partial_\mu \eta$.  We will gauge fix by choosing the Coulomb gauge, which  is not covariant, but since we will be interested in the non-relativistic case, this is of no consequence. To do this we include a term in the action 
\begin{equation}
S_{GF} = \frac{1}{2\zeta} \int  d^4x (\partial^{i}b_{i\nu}^I)^2
\end{equation}
where in the limit $\zeta\rightarrow 0$  the fields $A$ and $B$  become  longitudinal and transverse respectively obeying  $\vec{\nabla}\cdot \vec{A}^I=0$ and $\vec \nabla\times \vec{B}^I=0$ respectively.  These correspond to nine conditions
that reduces the number of degrees of freedom down to three. The interactions in the bulk follow in a similar fashion.


   To compute the propagators, we closely follow \cite{zaanen} and perform a coordinate transformation where the modes are decomposed into longitudinal component (L) and transverse components (R, S) with respect to the spatial momenta. 
   The transformation matrix on a vector is given by
\bea
 \begin{pmatrix}
 V_L \\
 V_R \\
 V_S \\
 \end{pmatrix}=
 \begin{pmatrix}
 -i\frac{q_x }{q} & -i\frac{q_y}{q} & -i\frac{q_z}{q} \\
 -\frac{q_{\perp}}{q} & \frac{q_x q_y}{q_\perp q} & \frac{q_x q_z}{q_\perp q} \\
 0 & -i\frac{q_z}{q_\perp} & i\frac{q_x}{q_\perp} \\
 \end{pmatrix} \begin{pmatrix}
 V_x \\
 V_y \\
 V_z \\
 \end{pmatrix}
\eea
where $q_\perp=\sqrt{q_y^2+q_z^2}$. One can now rewrite the leading order Lagrangian in this basis as
\bea
\label{eq:twopointaction}
 L^{LO}= \Phi^I_J \: K^{IJKL} \: \Phi^K_L 
\eea
where $\Phi^I_J=(A^I_J,B^I_J)$. The elastic moduli tensor in an isotropic lattice is singular due to the absence of antisymmetric strains and hence the kernel $K_{IJKL}$ is non-invertible. To invert the tensor, one can add the antisymmetric tensor 
\bea A_{IJKL}=\frac{1}{2}\left(\delta_{IK} \delta_{JL}-\delta_{IL} \delta_{JK}\right)\eea
to the elastic moduli tensor and set its contribution to zero at the end of the calculation. This allows us to compute the two-point correlators in this basis, where the kernel $K_{IJKL}$ becomes block diagonal. We obtain three transverse sectors 
\bea
\langle A_R^L A_R^L\rangle  &=& \frac{2c_T^2\omega^2}{k^2(c_T^2k^2-\omega^2)}  \nn \\
\langle A_R^L B_L^S\rangle  &=& \frac{2ic_T^2\omega}{k(c_T^2k^2-\omega^2)}  \nn \\
\langle  B_L^SB_L^S\rangle  &=&\frac{2c_T^2}{c_T^2k^2-\omega^2},
\nn \\
\eea

\bea
\langle  A_S^L  A_S^L\rangle  &=&\frac{2c_T^2\omega^2}{k^2(c_T^2k^2-\omega^2)}\nn \\
\langle  A_S^L B_L^R\rangle  &=&\frac{2ic_T^2\omega}{k(c_T^2k^2-\omega^2)}\nn \\
\langle   B_L^R B_L^R\rangle  &=&\frac{2c_T^2}{c_T^2k^2-\omega^2}, \nn \\
\eea

\bea
\langle A_R^R A_R^R\rangle  &=&\frac{-2}{k^2}=\langle A_R^R A_S^S\rangle  \nn \\
\langle A_R^S A_R^S\rangle  &=&\frac{2c_L^2\omega^2+8(c_T^2-c_L^2)c_T^2k^2}{k^2(c_L^2k^2-\omega^2)}=\langle A_S^R A_S^R\rangle,
\nn \\
\eea
and one longitudinal sector
\bea
\langle A_R^S A_S^R\rangle  &=&\frac{2(c_L^2-2c_T^2)(2c_T^2k^2-\omega^2)}{k^2(c_L^2k^2-\omega^2)}\nn \\
\langle A_R^S B_L^L\rangle  &=&\frac{-2i(c_L^2-2c_T^2)\omega}{k(c_L^2k^2-\omega^2)}=-\langle A_S^R B_L^L\rangle  \nn \\
\langle  B_L^L B_L^L\rangle  &=&\frac{2c_L^2}{c_L^2k^2-\omega^2},
\nn \\
\eea
where we have traded $\tilde{\kappa}$, $\mu$ for the speed of the longitudinal and transverse phonon modes $c_L, c_T$ via the relations
\bea
\mu &=& m \bar n \: c_T^2\nn\\ \tilde\kappa &=& m \bar n \: (c_L^2-\frac{4}{3}c_T^2)
\eea
Transforming back to the cartesian coordinates and plugging the above correlators in \eqref{eq:correctionPreInt} allows one to compute the corrections to the dislon action.
A generic term in the integrand has the following functional form: 
\bea\label{eq:polyterm}
I \subset \int d^2 k_{\perp} \frac{\omega^jk_x^mk_y^nk_z^l} {k^4(k^2-\frac{\omega^2}{c_T^2})(k^2-\frac{\omega^2}{c_L^2})} \eea
with $j+m+n+l = 8$ and $k_\perp=(k_x,k_y)$. We perform the integrals within dimensional regularization by analytically continuing to $d = 2-2\epsilon$ . 
Note that all terms with either $m$ or $n$ odd vanishes after integration. The contribution from the non-zero terms is given by 
\bea
I \subset \frac{1}{2}\frac{\Omega_{2-2\epsilon}}{(2\pi)^{2-2\epsilon}}\Gamma[1+\frac{p}{2}-\epsilon] \Gamma[3-\frac{p}{2}+\epsilon]\mu^{\epsilon}\nn\\
\omega^j k_z^l \iint dxdy(1-x)(1-y)^2\Delta^{-3+\frac{p}{2}-\epsilon} 
\eea
with $m+n=p$ and $\Delta = (k_z^2-xc_T^{-2}\omega^2-x(1-y)c_L^{-2}\omega^2)$.
\newpage
\bibliography{main.bib}

\end{document}